\documentclass[12pt]{article}
\usepackage{amssymb}
\usepackage{color,graphicx}
\usepackage{amsmath}

\usepackage{cite}
\usepackage{float}

\newcommand{\p}{\bot}
\newcommand{\dd}{\partial}
\newcommand{\de}{\delta}

\newcommand{\e}{\varepsilon}
\newcommand{\ls}{\left(}
\newcommand{\rs}{\right)}

\newcommand{\La}{\Lambda}
\newcommand{\m}{\mu}
\newcommand{\n}{\nu}

\newcommand{\De}{\Delta}

\newcommand{\disn}[2]{$$\displaylines{\refstepcounter{equation}\label{#1}\hskip 1em minus 1em #2\hfilneg}$$}
\newcommand{\nom}{\hfil\hskip 1em minus 1em (\theequation)}
\newcommand{\ns}{\hfill\cr\hfill}
\newcommand{\no}{\hfil \hskip 1em minus 1em\phantom{(\theequation)}
            \hfilneg\cr\hfilneg\hskip 1em minus 1em\hfil}

\begin{document}

\title{Construction of Perturbatively Correct Light Front Hamiltonian for (2+1)-Dimensional
Gauge Theory}

\author{
M.Yu.~Malyshev\thanks{M. Yu.~Malyshev works also in Petersburg Nuclear Physics Institute,
Gatchina (Saint Petersburg), e-mail: mimalysh@yandex.ru, m.malyshev@spbu.ru},
E.V.~Prokhvatilov\thanks{E-mail: e.prokhvatilov@spbu.ru},
R.A.~Zubov\thanks{E-mail: roman.zubov@hep.phys.spbu.ru},
V.A.~Franke\thanks{E-mail: v.franke@spbu.ru}\\
{\it Saint Petersburg State University, St. Petersburg, Russia}\\
}
\date{\vskip 15mm}
\maketitle

\begin{abstract}
In this paper we consider (2+1)-dimensional SU(N)-symmetric gauge theory within light front
perturbation theory, regularized  by the method analogous
to Pauli-Villars regularization. This enables us
to construct correct renormalized light front Hamiltonian.
\end{abstract}


\maketitle

\section{Introduction}
The present paper is a continuation of the previously initiated investigation of the possibility
to construct correct renormalized light front (LF) Hamiltonian
\cite{tmf97,tmf99,tmf02,
Paston.Prokhvatilov.Franke.Nucl.Phys.B.Proc.Suppl.2002,
Brodsky.Franke.Hiller.McCartor.Paston.Prokhvatilov2004,
Yad.Fiz.2005,NPPF,
Malyshev.Paston.Prokhvatilov.Zubov.IntJTheorPhys2015,
Malyshev.Paston.Prokhvatilov.Zubov.Franke.TMPh2015}.
Light front coordinates are defined as follows
\cite{dir}:
$$x^\pm=(x^0\pm x^1)/\sqrt{2}, \quad x^\p,$$
where $x^0$, $x^1$, $\,x^\p$ are Lorentz coordinates.
The Hamiltonian is obtained by canonical quantization on the
LF surface, i.e. $x^+=0$, where the $x^+$ plays the role of time.  This quantization leads to the
appearance of the singularity at zero value of the momentum $p_-$. So one has to regularize this singularity. The regularization can be introduced as the cutoff $|p_-|\geqslant\e$ in field modes.
However
the results of such regularized theory can turn out to be non-equivalent to results obtained with usual
quantization on the surface $x^0=const$ in Lorentz coordinates.
Indeed, such non-equivalence  is found  in perturbation theory framework when perturbation theory in coupling constant, generated by canonical Hamiltonian on the LF, and usual perturbation theory, corresponding to the quantization on
the instant time surface in Lorentz coordinates, are compared \cite{burlang,Burkardt.Langnau, tmf97}.
Beside of that, the  regularization $|p_-|\geqslant\e$ does not allow  correct
description of non-perturbative vacuum effects
\cite{Prokhvatilov.Franke.Phys.Atom.Nucl.1989,Ilgen}.
Let us also mention the approach  to the
construction of
effective LF  Hamiltonian, proposed  for quantum chromodynamics (QCD) in papers
\cite{Stanislaw.Glazek.and.Maria.Gomez.Rocha.arXiv.2015,
Glazek.Wilson.Phys.Rev.D1993,Glazek.Wilson.Phys.Rev.D1994}.
The above mentioned perturbative non-equivalence can be, in a principle, removed via introduction of special extra fields \cite{tmf97,tmf99}
playing the role
analogous to the role of extra fields in Pauli-Villars (PV) regularization \cite{PV}.
The introduction of these fields violates gauge invariance.
So the corresponding ultraviolet (UV) renormalization, which can restore this symmetry in the
regularization removing limit, was proposed.
In this way the renormalized LF Hamiltonian,
perturbatively equivalent to the Hamiltonian of the theory quantized on the instant time surface
in Lorentz coordinates, can be constructed.
However the coefficients of renormalization
counterterms in this Hamiltonian remain unknown parameters because their values are actually defined by the
sum of   infinite number  Feynman diagrams.
Nevertheless, if one considers, instead of
(3+1)-dimensional QCD, its (2+1)-dimensional analog, the number of such diagrams becomes finite
(due to superrenormalizability) and one can find explicit expressions for these coefficients in terms of original parameters of the theory including the regularization
parameters.
Such a model for pure gauge fields was considered in paper \cite{Malyshev.Paston.Prokhvatilov.Zubov.Franke.TMPh2015}.
However the analysis of perturbation theory diagrams was expounded
%(outlined)
rather briefly.
Besides the expression for the renormalized LF Hamiltonian also was not obtained.
In the present paper we give the details for
the evidence of different statements in the paper \cite{Malyshev.Paston.Prokhvatilov.Zubov.Franke.TMPh2015}  and  construct the renormalized LF Hamiltonian.

\section{Possible differences between perturbation theory on the LF and one on the instant time surface
in Lorentz coordinates}
We consider, like in \cite{Malyshev.Paston.Prokhvatilov.Zubov.Franke.TMPh2015}, the Yang-Mills theory in (2+1)-dimensional space-time.
Diagrams of
covariant perturbation theory for this model can differ for the quantization  on the surface of
constant time in Lorentz coordinates and for the quantization on the LF, depending on the way of the regularization.

We compare such diagrams using the following basic statement proved in the paper \cite{tmf97}: denote the propagator momenta by $Q_\m=(Q_+, Q_-, Q_\p)$, where $Q_\pm=(Q_0\pm Q_1)/\sqrt{2}$ are the momenta in LF coordinates and $Q_0, Q_1, Q_\p$ are the momenta in Lorentz coordinates.
Then the expression for a diagram in Lorentz coordinates can be written as the following integral over the momenta in LF coordinates (for brevity we drop the integration in $Q_\p$) \cite{tmf97}:
\disn{1}{
\prod_i \int dQ_{i+} \int_B dQ_{i-} f(Q_i, p_j),
\nom}
where $f(Q_i, p_j)$ denotes the integrand which includes delta-functions expressing momentum conservation in vertices. The integration domain in $Q_{i-}$ is limited by the quantity $B$ which has the order of maximal external momentum $p_{j-}$ (i.e. it is bounded by the domain of the size $s\,|p_{i-}^{max}|$, where the constant $s$ depends on the diagram structure).
If we introduce in this integral also the cutoff $|Q_{i-}|\geqslant\e$ we get the result corresponding to the diagram, calculated in LF coordinates where such cutoff is used conventionally.
The limitation of the integration over $Q_{i-}$ by the domain $B$ can be related to analogous limitation in old fashioned LF perturbation theory where the momenta $Q_{i-}\geqslant0$ in the intermediate states do not exceed the total momentum.
Thus we obtain that the difference between results of calculation of a diagram in Lorentz and LF coordinates is given by the sum of contributions to the integral (\ref{1}) from domains with $|Q_{i-}|\leqslant\e$.
In the following the different terms of this sum we call configurations as in \cite{tmf97,NPPF}.

Let us consider (2+1)-dimensional Yang-Mills theory with the following Lagrangian density:
\disn{2}{
{\cal L} =- \frac{1}{4} F_{\m\n}^a F^{a\m\n} + \frac{m}{2} \e^{\m\n\alpha} \ls A_{\m}^a \partial_{\n} A_{\alpha}^a  +
\frac{2}{3} g f^{abc}A_{\m}^a A_{\n}^b A_{\alpha}^c \rs,
\nom}
where $A_{\m}^a(x)$ is the $SU(N)$ gauge field,
$F_{\m\n}^a=\dd_{\m}A_{\n}^a - \dd_{\n} A_{\m}^a+gf_{abc}A_{\m}^bA_{\n}^c$,
$a=1,...,N^2-1$ are adjoint representation indices, $f_{abc}$ are structure constants, $m$ is mass, $g$ is coupling constant and $\e^{\m\n\alpha}$ is the Levi-Civita symbol.
We add Chern-Simons term to regularize infrared (IR) divergences \cite{Deser.Jackiw.Templeton.Annals.of.Physics.2000,
Deser.Jackiw.Templeton.Phys.Rev.Lett.1982}.

Further we choose LF gauge $A_-=0$. This leads to the disappearance of the term of power four in fields $A_{\m}$ in eq-n (\ref{2}).

It was noticed in the previous paper \cite{Malyshev.Paston.Prokhvatilov.Zubov.Franke.TMPh2015} that only finite number of UV divergent diagrams are to be renormalized in this model (these diagrams are shown explicitly in \cite{Malyshev.Paston.Prokhvatilov.Zubov.Franke.TMPh2015}, indices of UV divergency of them do not exceed one).

Let us consider the variant of UV regularization of this theory in which we introduce both extra fields and higher derivatives.
We take into account that the chosen gauge $A_-=0$ generates spurious singularity in the gluon propagator (namely, additional pole in $p_-$).
To regularize this singularity we introduce one extra field more and new parameter $\m$ \cite{Malyshev.Paston.Prokhvatilov.Zubov.Franke.TMPh2015}.
We start  from the theory with only one extra field and show
that in this theory we have the differences between calculations of diagrams of perturbation theory,
generated by the  quantization  on the constant time surface, $x^0=0$, and generated by the  quantization  on the LF, $x^+=0$.
The Lagrangian has the following form:
\disn{3}{
{\cal L}=-\frac{1}{4} f_1^{a\m\n}\ls\frac{\La^2+2\partial_+\partial_-}{\La^2}\rs f_{1,\m\n}^a+
\frac{m}{2} \e^{\m\n\alpha} A_{1,\m}^a\ls\frac{\La^2+2\partial_+\partial_-}{\La^2}\rs\partial_\n A_{1,\alpha}^a+\ns+\frac{1}{4} f_2^{a\m\n}\ls\frac{\m^2+2\partial_+\partial_-}{\m^2}\rs f_{2,\m\n}^a+
\frac{m}{2} \e^{\m\n\alpha} A_{2,\m}^a\ls\frac{\m^2+2\partial_+\partial_-}{\m^2}\rs\partial_\n A_{2,\alpha}^a+\ns+
g f^{abc}A_{\m}^a A_{\n}^b \dd^\m A^{c\n},
\nom}
where $f_{j,\m\n}^a=\dd_\m A_{j,\n}^a-\dd_\n A_{j,\m}^a$, $j=1,2$, $A_{1,\m}^a$ is the original field, $A_{2,\m}^a$ is the extra field. The interaction term contains the sum of these fields,
$A_{\m}^a=A_{1,\m}^a+A_{2,\m}^a$. Beside of that, we require $A_{j,-}^a=0$.

The propagator of the field $A_{j\m}$ has the following form:
\disn{4}{
\Delta_{1,\m\n}^{ab}=
\frac{i\delta^{ab}}{Q^2-m^2+i0}
\ls g_{\m\n}-\frac{Q_{\m}n_{\n}+n_{\m}Q_{\n}+
i\,m\,\e_{\m\n\alpha}n^{\alpha}}{2Q_+Q_-+i0}2Q_+\rs
\frac{1}{\frac{2Q_+Q_-}{\La^2}-1+i0},\ns
\Delta_{2,\m\n}^{ab}=
\frac{-i\delta^{ab}}{Q^2-m^2+i0}
\ls g_{\m\n}-\frac{Q_{\m}n_{\n}+n_{\m}Q_{\n}+
i\,m\,\e_{\m\n\alpha}n^{\alpha}}{2Q_+Q_-+i0}2Q_+\rs
\frac{1}{\frac{2Q_+Q_-}{\m^2}-1+i0},
\nom}
where the vector $n_\m$ has components $n_+=1$, $n_-=n_{\p}=0$, $n_\m n^\m=0$.

The analysis of the UV divergences in the perturbation theory with this Lagrangian shows that the introduced regularization is sufficient for the removing of  these divergences  \cite{Malyshev.Paston.Prokhvatilov.Zubov.Franke.TMPh2015}.
Since the interaction terms depend only on the sum of fields, $A_\m=A_{1,\m}+A_{2,\m}$, the perturbation theory can be formulated with the summary propagator:
\disn{4a}{
\Delta_{\m\n}^{ab}=\Delta_{1,\m\n}^{ab}+\Delta_{2,\m\n}^{ab}=
\frac{-i\delta^{ab}}{Q^2-m^2+i0}
\ls g_{\m\n}-\frac{Q_{\m}n_{\n}+n_{\m}Q_{\n}+
i\,m\,\e_{\m\n\alpha}n^{\alpha}}{2Q_+Q_-+i0}2Q_+\rs R, \ns
R=\frac{-1}{\frac{2Q_+Q_-}{\La^2}-1+i0}+
\frac{1}{\frac{2Q_+Q_-}{\m^2}-1+i0}=
\frac{2Q_+Q_-\ls\m^2-\La^2\rs}{\ls 2Q_+Q_--\m^2+i0\rs\ls 2Q_+Q_--\La^2+i0\rs}.
\nom}
Note that the regularization is removed in the limit $\m\to0$, $\La\to\infty$ ($R\to1$).

Let us write the expressions for the propagator at different Lorentz indices (for the analysis below we do not use the SU(N)-indices  because they are not essential  there):
\disn{5}{
\De_{++}(Q)=4Q_+^2\De(Q), \quad
\De_{+\p}(Q)=2Q_+(Q_\p+im)\De(Q), \quad
\De_{\p\p}(Q)=2Q_+Q_-\De(Q), \ns \text{where} \quad \De(Q)=-i(\La^2-\m^2)\prod_{l=0}^2\ls 2Q_+Q_--m_l^2+i0\rs^{-1},
\nom}
$m_0^2=m^2+Q_\p^2$, $m_1=\La$, $m_2=\m$.

\begin{figure}[h!]
  \centering
  \includegraphics[width=160mm]{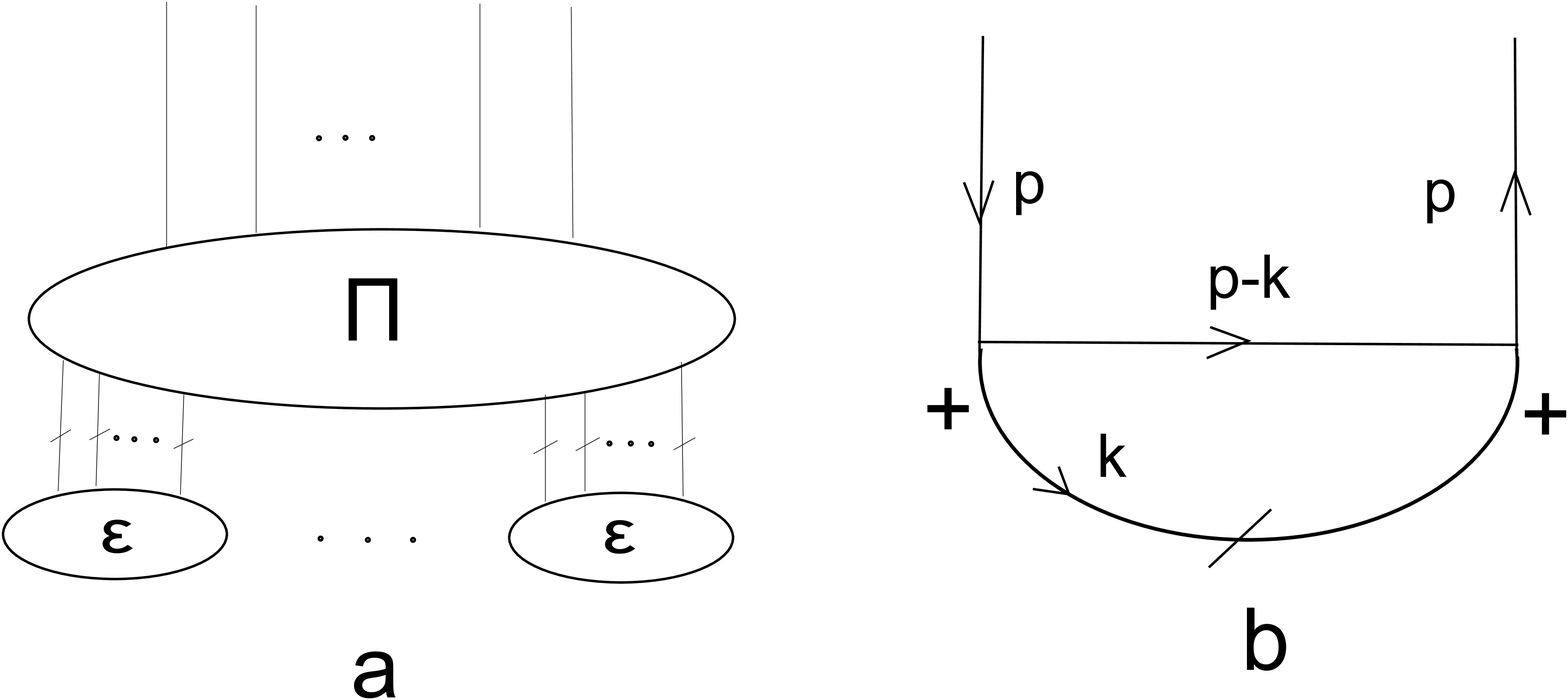}
  \caption{the most common kind of configuration (a). The simplest configuration example (b).}
\label{Fig1}
\end{figure}

Let us remind that  possible differences between calculations of Feynman diagrams on the LF and  in Lorentz  coordinates are related to the contribution to the corresponding Feynman diagrams from the integration domains $|Q_{i-}|\leqslant\e$, which are absent in diagrams of the theory on the LF with the regularization $|Q_{i-}|\geqslant\e$.
Let us show that the configurations contributing to  such  difference have the form shown in Fig.~\ref{Fig1}а, where the crossed lines denote the propagators which have momenta limited by the condition $|Q_{i-}|\leqslant \e$.

Here we have extracted the "$\e$-blocks"{} which include both the set of  joined crossed lines and also the  not crossed lines if the momenta $Q_{-}$ of the latter are limited by the quantity of order $\e$  (further  we call these lines $\Pi$-lines; for example, one gets these lines  if they are joined only with crossed lines and then their momenta are limited by the domain $B$ of order $\e$ by the statement in the beginning of the present section).
Besides we have extracted the $\Pi$-block which contains all  remaining $\Pi$-lines. All external lines are joined to this $\Pi$-block. Let us begin with a simplest example of such a configuration (Fig.~\ref{Fig1}b). The corresponding Feynman integral has the following form:
\disn{6}{
I(p)=-4g^2N\ls\La^2-\m^2\rs^2\int dk_+ \int_{-\e}^\e dk_-\frac{(2p_--k_-)^2(p_+-k_+)(p_--k_-)}{(\prod_{l=0}^2\ls 2(p_+-k_+)(p_--k_-)-m_l^2+i0\rs)}\times\ns
\times\frac{k_+^2}{\prod_{l=0}^2\ls 2k_+k_--m_l^2+i0\rs}.
\nom}
To estimate this integral in the $\e\to0$ limit we remove the dependence on $\e$ of integration limits in $k_-$ by the change $k_-\to k_- \e$  and additionally $k_+\to k_+/\e$. Then  the integral takes the form:
\disn{7}{
-8g^2N\ls\La^2-\m^2\rs^2\int dk_+ \int_{-1}^1 dk_-\frac{(2p_--\e k_-)^2\,(p_+-k_+/\e)(p_--\e k_-)}{\prod_{l=0}^2\ls2(p_+-k_+/\e)(p_--\e k_-)-m_l^2+i0\rs}\times\ns
\times\frac{k_+^2/\e^2}{\prod_{l=0}^2\ls 2k_+k_--m_l^2+i0\rs}=\ns=
-8g^2N\ls\La^2-\m^2\rs^2\int dk_+ \int_{-1}^1 dk_-\frac{(2p_--\e k_-)^2\,(\e p_+-k_+)(p_--\e k_-)}{\prod_{l=0}^2\ls2(\e p_+-k_+)(p_--\e k_-)-\e m_l^2+i0\rs}\times\ns
\times\frac{k_+^2}{\prod_{l=0}^2\ls 2k_+k_--m_l^2+i0\rs}.
\nom}
Since the integral over $k_+$ is convergent we get the following finite expression in the $\e\to0$ limit:
\disn{8}{
-4g^2N\ls\La^2-\m^2\rs^2\int dk_+ \int_{-1}^1 dk_-\frac{1}{\prod_{l=0}^2\ls 2k_+k_--m_l^2+i0\rs},
\nom}
i.e. we have the difference between the calculation of such diagram on the LF  and its conventional calculation   in Lorentz coordinates.
To estimate  the dependence on the $\e$ in  general form we  consider at first how the propagators change under the above mentioned change of  integration variables.
Consider the case  when the propagator momentum $ Q_-$ is limited,  $|Q_-|\leqslant\e$. Let us write the $k$ instead of $Q$ (as in the Fig.~\ref{Fig1}b)  so that $|k_-|\leqslant\e$. Let us change $k_+\to k_+'=\e k_+$, $k_-\to k_-'= k_-/\e$. Then, in accordance with the eq-n (\ref{5}), we get  the following for the components of the propagator:
\disn{9}{\De(k) \to \De'(k')=\De(k')\sim O(1), \no
\De_{++}(k)\to\De_{++}'(k')=\De_{++}(k')/\e^2,\no
\De_{+\p}(k)\to\De_{+\p}'(k')=\De_{+\p}(k')/\e,\no
\De_{\p\p}(k)\to\De_{\p\p}'(k')=\De_{\p\p}(k')\sim O(1).
\nom}
Consider the case when the propagator momentum is $Q_\m=p_\m-k_\m$ where $p_-$ is finite and $|k_-|\leqslant\e$. Then we have the following:
\disn{10}{
\De(p-k) \to\De'(p,k')=\frac{-i\,\ls m_1^2-m_2^2\rs\e^3}{(-p_-+\e k_-')^3\prod_{l=0}^2(2k_+'-2\e p_++\e m_l^2/(p_--\e k_-')+i0/p_-)}\sim O(\e^3), \no
\De_{++}(p-k)\to\frac{4(\e p_+-k'_+)^2\De'(p,k')}{\e^2}\sim O(\e), \no
\De_{+\p}(p-k) \to\frac{2(\e p_+-k'_+)(p_\p-k_\p+im)\De'(p,k')}{\e}\sim O(\e^2), \no
\De_{\p\p}(p-k) \to\frac{2(\e p_+-k'_+)(p_--\e k'_-)\De'(p,k')}{\e}\sim O(\e^2).
\nom}
All vertex factors can be estimated by  finite quantity except of those cases when three $\e$-lines converge in the vertex (then they can be estimated by the quantity of order~$\e$).
Using these estimations one can easily prove that the diagram in the Fig.~\ref{Fig1}b tends to zero in $\e\to0$ limit at any other choice of indices.
An addition of more $\e$-lines to the $\e$-line,  i. e. the formation of the $\e$-block,  does not change the estimation of the diagram in Fig.~\ref{Fig1}b.
This can be shown in the following way: the $\e$-block consists from vertices and propagators with   every vertex being proportional to $\e$ and having one index +  which leads to the appearance of the factor $k'_+/\e$ in the propagator adjoint to this vertex.
So  the dependence on $\e$ for this vertex is cancelled. This is  true for propagators having indices ++ and  giving the contribution proportional to $1/\e^2$  and  for propagators having indices $+\p$  and giving the contribution proportional to $1/\e$. The propagator  with the indices $\p\p$ gives the contribution of order $O(1)$.
The  same situation takes place for the other vertices of the $\e$-block.

\begin{figure}[h!]
  \centering
  \includegraphics[width=160mm]{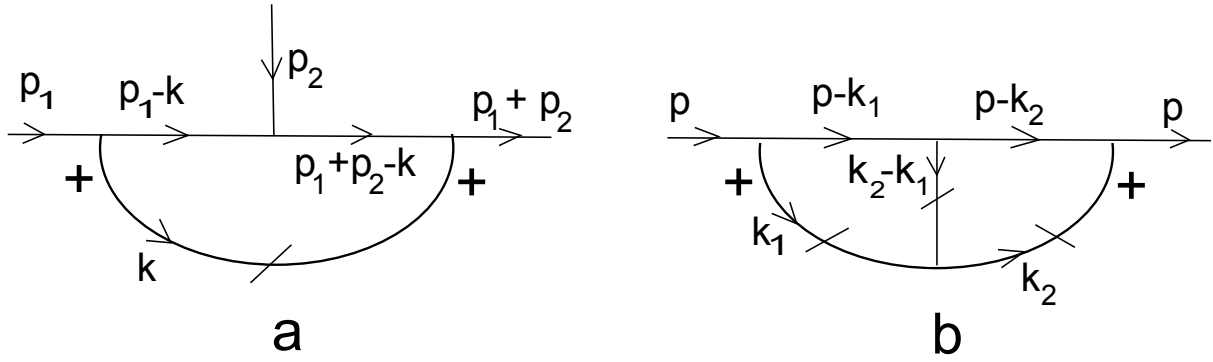}
  \caption{the complications of the simplest configuration.}
\label{Fig2}
\end{figure}

The next simple modification of the configuration in Fig.~\ref{Fig1}b has the form shown in Fig.~\ref{Fig2}a which differs from Fig.~\ref{Fig1}b by the addition of one external line.
Such diagram has two $\Pi$-lines through which the momentum $k$ of $\e$-lines goes. So, with the estimations of eq-ns (\ref{9}) and (\ref{10}) we obtain that  the configuration tends to zero as $\e\to0$.
The analogous situation has place when we increase the number   of $\Pi$-lines, through which the momentum $k$ of $\e$-lines goes, e.g. when we increase the number of external lines.

Let us   consider another case of configuration changing for Fig.~\ref{Fig1}b when we add one $\e$-line connecting $\Pi$- and $\e$-lines of the original configuration (Fig.~\ref{Fig2}b).
In this case  the $\Pi$- and $\e$-lines of the original configuration are divided into two parts where the  momenta $k_1$ and  $k_2$ of propagators of the divided $\e$-line can be chosen as loop integration variables so  that they go through one of the propagator of the divided $\Pi$-line (see Fig.~\ref{Fig2}b).
Then the estimations, corresponding to eq-ns (\ref{9}) and (\ref{10}),   and taking into account  the contribution of the new vertex adjoint to the original $\e$-line lead in result  that the configuration goes to zero as $\e\to0$.

Thus the most general form of the non equal to zero configuration can be pictured in analogy with the diagram (Fig.~\ref{Fig1}b)  in which the crossed line is changed to arbitrary $\e$-block.

So we come to the conclusion that in the theory under consideration we have nonzero configurations,  and  all of them are finite in the limit $\e\to0$.

Let us show now  that the method of the paper \cite{Malyshev.Paston.Prokhvatilov.Zubov.Franke.TMPh2015}  where  one more extra field is to be added allows to get going to zero all nonzero configurations considered above.  The Lagrangian in this case has the following form:
\disn{12}{
{\cal L}=\sum_{l=1}^3\ls -\frac{r_l}{4} f_l^{a\m\n}\ls\frac{m_l^2+2\partial_+\partial_-}{M_l^2}\rs f_{l,\m\n}^a+
r_l\frac{m}{2} \e^{\m\n\alpha} A_{l,\m}^a\ls\frac{m_l^2+2\partial_+\partial_-}{M_l^2}\rs\partial_\n A_{l,\alpha}^a\rs+\ns+
g f^{abc}A_{\m}^a A_{\n}^b \dd^\m A^{c\n},
\nom}
where $A_{1,\m}^a$ is the original field, $A_{2,\m}^a$
and $A_{3,\m}^a$ are the extra fields, $A_{l,-}^a=0$, $A_{\m}^a=A_{1,\m}^a+A_{2,\m}^a+A_{3,\m}^a$, the parameters $m$, $\,m_1=\La$, $m_2=\m$ are analogous to the parameters in the Lagrangian (\ref{3}), $r_l=\pm1$.
The parameter $m_3$ is new parameter related to the introduction of the additional field $A_{3,\m}^a$.
The quantities $M_l^2$ are to be expressed in terms of $m_l$ and $m$ so that the index of UV divergency of the summary propagator were by  two units less. This summary propagator is the sum of each field propagators:
\disn{12a}{
\Delta^{ab}_{\m\n}=
\frac{-i\delta^{ab}}{\ls k^2-m^2+i0\rs}
\ls g_{\m\n}-\frac{k_{\m}n_{\n}+n_{\m}k_{\n}+
i\,m\,\e_{\m\n\alpha}n^{\alpha}}{2k_+k_-+i0}2k_+\rs
\ls\sum_{l=1}^3\frac{r_lM_l^2}{m_l^2-2k_+k_--i0}\rs.
\nom}
If we reduce the sum in eq. (\ref{12a}) to common denominator and require the disappearance of the terms  of power four and zero in $k$ in the numerator of this sum we obtain the following conditions:
\disn{12aa}{
\sum_{l=1}^3r_lM_l^2=0, \quad \sum_{l=1}^3\frac{r_lM_l^2}{m_l^2}=0.
\nom}
Further we require  that the obtained expression for the summary propagator  must go to the expression for the propagator of unregulated theory in the limit $m_2^2=\m^2\to0$, $m_1^2=\La^2\to\infty$, $m_3^2\to\infty$, i.e.
\disn{12b}{
2k_+k_-\frac{r_1M_1^2(m_2^2+m_3^2)+r_2M_2^2(m_1^2+m_3^2)+
r_3M_3^2(m_1^2+m_2^2)}{\prod_{l=1}^3(2k_+k_--m_l^2+i0)}\to\ns\to
\frac{r_1M_1^2m_3^2+r_2M_2^2(m_1^2+m_3^2)+
r_3M_3^2m_1^2}{m_1^2m_3^2}=\ns=
 r_1\frac{M_1^2}{m_1^2}
+r_2\ls\frac{1}{m_1^2}+\frac{1}{m_3^2}\rs M_2^2+r_3\frac{M_3^2}{m_3^2}\to1.
\nom}
If we take  into account the second condition in eq. (\ref{12aa}),  we can satisfy the condition of eq. (\ref{12b}) taking,  for example,  $\frac{r_2M_2^2}{m_2^2}=-1$. In accordance with the eq. (\ref{12aa}) we obtain:
\disn{13}{
r_1M_1^2=\frac{m_1^2(m_3^2-m_2^2)}{m_3^2-m_1^2},\quad
\no
r_3M_3^2=-\frac{m_3^2(m_1^2-m_2^2)}{m_3^2-m_1^2}.
\nom}
Beside of that we require that only the field $A_1$ be remained in the regularization
removing limit $m_2^2\to0$, $m_1^2\to\infty$, $m_3^2\to\infty$.  With this aim we require additionally $\frac{m_3^2}{m_1^2}\to\infty$. We see from these facts that $r_1=1$, $r_2=-1$, $r_3=-1$.

To estimate the dependence of different configurations on $\e$ let us rewrite the summary propagator in the following form:
\disn{15}{
\Delta_{\m\n}^{ab}(Q)=
\frac{i\delta^{ab}\ls 2Q_+Q_-g_{\m\n}-2Q_+(Q_{\m}n_{\n}+Q_{\n}n_{\m}+
i\,m\,\e_{\m\n\alpha}n^{\alpha})\rs}{\prod_{l=0}^3\ls 2Q_+Q_--m_l^2+i0\rs}R,
\nom}
where
$$m_0^2=m^2+Q_\p^2,\quad \text{а} \quad R=r_1M_1^2(m_2^2+m_3^2)+r_2M_2^2(m_1^2+m_3^2)+
r_3M_3^2(m_1^2+m_2^2).$$

The  corresponding to the eq.~(\ref{9}) estimations for the propagator (\ref{15})  keep its form, and the estimations,  corresponding to the eq.~(\ref{10}), take the following form:
\disn{15a}{
\De'(p,k')\sim O(\e^4), \no
\De_{++}'(p,k')\sim O(\e^2), \no
\De_{+\p}'(p,k')\sim O(\e^3), \no
\De_{\p\p}'(p,k')\sim O(\e^3),
\nom}
i.e. the power of $\e$ increases comparatively to eq-ns (\ref{10}). This leads to disappearance of all before considered configurations in the limit $\e\to0$.
Thus in the theory with the Lagrangian (\ref{12}) we get the coincidence of diagram values  in    Lorentz coordinate  instant time  quantization  and LF quantization.

Using the results of the paper \cite{tmf99} we have shown in the paper \cite{Malyshev.Paston.Prokhvatilov.Zubov.Franke.TMPh2015}  that it is possible to avoid related to parameter $\m$ IR divergences in the  regularization removing limit if one satisfies the condition $(\ln{\m})^n/\La\to0$ in the limit $\m\to0$, $\La\to\infty$ where $n$ is arbitrary integer number (e.g. one can take $\m=m^2/\La$).

\begin{figure}[h!]
  \centering
  \includegraphics[width=160mm]{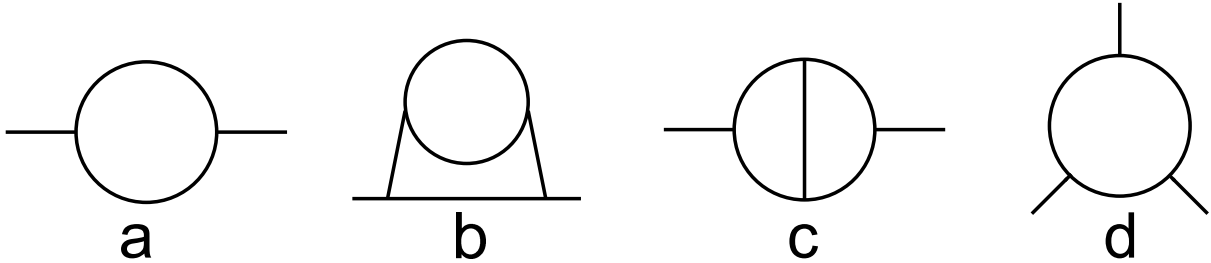}
  \caption{the diagrams divergent in the UV regularization removing limit.}
\label{Fig3}
\end{figure}

Owing to superrenormalizability of considered theory in (2+1)-dimensional space we need to analyze and calculate only finite number of UV divergent diagrams  shown in Fig.~\ref{Fig3}.
It was proved in the paper \cite{Malyshev.Paston.Prokhvatilov.Zubov.Franke.TMPh2015}  that the 1-loop diagram with two external lines (Fig.~\ref{Fig3}a), having the  index of UV divergency equal to unity, and the 1-loop diagram with three external lines and with the index of UV divergency equal to zero (Fig.~\ref{Fig3}d) turn out to be finite after exact analytic calculation.
The remaining 2-loop diagrams with two external lines (Fig.~\ref{Fig3}b, \ref{Fig3}c), having the index of UV divergency equal to zero, can be renormalized by the comparison with corresponding renormalized dimensionally regularized diagrams.
The renormalization counterterms are found at that  by the calculation of the difference of the expressions for these diagrams in the considered  and dimensional regularizations.
Such a procedure of the renormalization allows to restore gauge and Lorentz symmetry in the theory with the Lagrangian (\ref{12}) in the regularization removing limit.
In result we have to add to the Lagrangian (\ref{12}) the counterterm of the form $C\,\ln{\ls\La/m\rs}A_\p^aA_\p^a$ where the constant $C$ can be found as a result of numerical calculations of diagrams (Fig.~\ref{Fig3}b, \ref{Fig3}c).
This form of the counterterm is in agreement with the remaining Lorentz invariance of the Lagrangian (\ref{12}) under the following coordinate transformation: $x^+\to\lambda x^+$, $x^-\to\lambda^{-1} x^-$, $x^\p\to x^\p$.
Due to this invariance the logarithmically divergent diagrams (Fig.~\ref{Fig3}b, \ref{Fig3}c) calculated at zero external momentum (and at external Lorentz indices excluding the "$-$"{}) are equal to zero if they have at least one  external Lorentz index "+"{}.
Thus the only diagrams that contribute to the counterterm are those  which have only external indices "$\p$"{}.  We give in the Appendix the expression for the diagram shown in Fig.~\ref{Fig3}a.

\section{The construction of the renormalized LF Hamiltonian}
Consider the theory with the Lagrangian (\ref{12}) and rewrite the expression for this Lagrangian in the gauge $A_{l,-}^a=0$.
\disn{14}{
{\cal L}=\sum_{l=1}^3r_l\Biggl(-\frac{\partial_+ f_{l,+-}^a
\partial_- f_{l,+-}^a}{M_l^2}+\frac{m_l^2}{2M_l^2}(f_{l,+-}^a)^2+
\partial_+A_{l,\p}^a\frac{2\partial_-^{2}}{M_l^2}
\partial_+A_{l,\p}^a+\frac{m_l^2}{M_l^2}\partial_+A_{l,\p}^a
\partial_-A_{l,\p}^a+\ns
+m\ls\frac{m_l^2}{M_l^2}f_{l,+-}^a
A_{l,\p}^a+\frac{2}{M_l^2}f_{l,+-}^a
\partial_+\partial_-A_{l,\p}^a\rs-\partial_\p f_{l,+-}^a
\ls\frac{m_l^2+2\partial_+\partial_-}{M_l^2}\rs A_{l,\p}^a\Biggr)-\ns
-gf^{abc}A_+^aA_\p^b\partial_-A_\p^c.
\nom}

To write the action corresponding to this Lagrangian in Hamiltonian form we find at first the momentum $\Pi_{l,\p}^a$ conjugated to the field $A_{l,\p}^a$:
\disn{16}{
\Pi_{l,\p}^a=\frac{\partial {\cal L}}{\partial(\partial_+A_{l,\p}^a)}=r_l\ls4\frac{\partial_-^2}
{M_l^2}\partial_+A_{l,\p}^a+\frac{m_l^2}{M_l^2}
\partial_-A_{l,\p}^a+\frac{2}{M_l^2}(\partial_\p-m)
\partial_-f_{l,+-}^a\rs.
\nom}
This relation allows to express the quantity $\partial_+ A_{l,\p}^a$ in terms of variables $\Pi_{l,\p}^a$, $A_{l,\p}^a$ and $f_{l,+-}^a$:
\disn{17}{
\partial_+A_{l,\p}^a=\frac{M_l^2}{4}\partial_-^{-2}\ls r_l\Pi_{l,\p}^a-\frac{m_l^2}{M_l^2}
\partial_-A_{l,\p}^a-\frac{2}{M_l^2}(\partial_\p-m)
\partial_-f_{l,+-}^a\rs.
\nom}
In result the action can be rewritten in the following form:
\disn{18}{
S=\int dx^+dx^-dx^\p\Bigg[\sum_{l=1}^3\ls-\frac{r_l}{M_l^2}\partial_+f_{l,+-}^a
\partial_-f_{l,+-}^a+\Pi_{l,\p}^a\partial_+ A_{l,\p}^a\rs-{\cal H}\Bigg],
\nom}
where
\disn{19}{
{\cal H}(\Pi_{l,\p}^a, A_{l,\p}^a, f_{l,+-}^a)=\sum_{l=1}^3\Bigg[-\frac{r_lm_l^2}{2M_l^2}f_{l,+-}^a
f_{l,+-}^a+\frac{r_lm_l^2}{M_l^2}A_{l,\p}^a
(\partial_\p-m)f_{l,+-}^a+\ns+\frac{r_lM_l^2}{8}\ls r_l\Pi_{l,\p}^a-\frac{m_l^2}{M_l^2}\partial_-A_{l,\p}^a-
\frac{2}{M_l^2}(\partial_\p-m)\partial_-f_{l,+-}^a\rs\times\ns\times \partial_-^{-2}\ls r_l\Pi_{l,\p}^a-\frac{m_l^2}{M_l^2}\partial_-A_{l,\p}^a-
\frac{2}{M_l^2}(\partial_\p-m)\partial_-f_{l,+-}^a\rs\Bigg]+\ns+gf^{abc}A_+^a
A_\p^b\partial_-A_\p^c-C\ln{\ls\La/m\rs}A_\p^aA_\p^a.
\nom}
The expression (\ref{18}) shows that the canonical momentum conjugated to the field $f_{l,+-}^a$ generates the canonical constraint on the LF which turns out to be the constraint of the second kind.
To avoid this constraint let us go to new variables with the help of Fourier transformation in $x^-$ (introducing the regularization $|p_-|\geqslant\varepsilon$):
\disn{20}{
f_{l,+-}^a(x)=\frac{M_l}{\sqrt{2\pi}}\int\limits_{(|p_-|>\e)} \frac{dp_-}{\sqrt{2|p_-|}}\bigg( a_l^a(p_-;x^\p,x^+)\exp{(-ip_-x^-)}\bigg)=\ns=
\frac{M_l}{\sqrt{2\pi}}\int\limits_\e^\infty \frac{dp_-}{\sqrt{2p_-}}\bigg( a_l^a(p_-;x^\p,x^+)\exp{(-ip_-x^-)}+
a_l^{a+}(p_-;x^\p,x^+)\exp{(ip_-x^-)}\bigg).
\nom}
Then the term in the action,
\disn{21}{
\sum_{l=1}^3\int dx^+dx^-dx^\p\ls-\frac{r_l}{M_l^2}\partial_+f_{l,+-}^a
\partial_-f_{l,+-}^a\rs
\nom}
can be   transformed as follows (after integration by parts in  $x^+$):
\disn{22}{
\sum_{l=1}^3\int dx^+dx^\p\int_\varepsilon^\infty dp_-\ls-ir_la_l^{a+}(p_-;x^\p,x^+)\partial_+a_l^a(p_-;x^\p,x^+)\rs.
\nom}
The obtained Hamiltonian form of the action allows to define quantum commutation relations  for canonical variables:
\disn{23}{
\Big[a_l^{a}(p_-;x^\p,x^+), a_{l'}^{b+}(p_-';x'^{\p},x^+)\Big]=-r_l\de_{ll'}
\de^{ab}\de(p_--p_-')\de(x^\p-x'^{\p}),
\nom}
\disn{24}{
\Big[A_{l,\p}^{a}(x), \Pi_{l',\p}^{b}(x')\Big]_{x'^+=x^+}=i\delta_{ll'}
\delta^{ab}\delta(x^--x'^-)\delta(x^\p-x'^\p).
\nom}
The remaining commutation relations  for these fields are equal to zero. The variables $a_l^{a}(p_-;x^\p,x^+)$ and $a_l^{a+}(p_-;x^\p,x^+)$ can be considered as that analogous to creation and annihilation operators on the LF.  Let us introduce also the Fourier transformation for the variables $A_{l,\p}^{a}(x)$ and $\Pi_{l,\p}^{b}(x)$:
\disn{25}{
A_{l,\p}^{a}(x)=\frac{M_l}{m_l\sqrt{2\pi}}\int\limits_{(|p_-|>\e)} \frac{dp_-}{\sqrt{2|p_-|}}\bigg(\bigg( a_{l,\p}^a(p_-;x^\p,x^+)+a_{l,\p}^{a+}(-p_-;x^\p,x^+)\bigg)\exp{(-ip_-x^-)}\bigg)=\ns=\frac{M_l}{m_l}\int_\e^\infty
\frac{dp_-}{\sqrt{4\pi p_-}}\Bigl[\Bigl(a_{l,\p}^{a}(p_-;x^\p,x^+)+
b_{l,\p}^{a}(p_-;x^\p,x^+)\Bigr)\exp(-ip_-x^-)+\ns+\Bigl(a_{l,\p}^{a+}(p_-;x^\p,x^+)+
b_{l,\p}^{a+}(p_-;x^\p,x^+)\Bigr)\exp(ip_-x^-)\Bigr],\no
\Pi_{l,\p}^{b}(x)=\frac{-im_l}{M_l\sqrt{2\pi}}\int\limits_{(|p_-|>\e)} dp_-\sqrt{\frac{|p_-|}{2}}\bigg(\bigg( a_{l,\p}^b(p_-;x^\p,x^+)-a_{l,\p}^{b+}(-p_-;x^\p,x^+)\bigg)\exp{(-ip_-x^-)}\bigg)=\ns
=\frac{-im_l}{M_l}\int_\e^\infty dp_-
\sqrt{\frac{p_-}{4\pi}}\Bigl[\Bigl(a_{l,\p}^{b}(p_-;x^\p,x^+)-
b_{l,\p}^{b}(p_-;x^\p,x^+)\Bigr)\exp(-ip_-x^-)-\text{h.c.}\Bigr],
\nom}
where we introduce the denotation
$$\,\,\,b_{l,\p}^{a}(p_-;x^\p,x^+)=a_{l,\p}^{a+}(-p_-;x^\p,x^+).$$
The operators $a_{l,\p}^{a}(p_-;x^\p,x^+)$, $a_{l,\p}^{a+}(p_-;x^\p,x^+)$, $b_{l,\p}^{a}(p_-;x^\p,x^+)$, $b_{l,\p}^{a+}(p_-;x^\p,x^+)$ satisfy the following canonical commutation relations analogous to commutation relations for the creation and annihilation operators:
\disn{26}{
\Big[a_{l,\p}^{a}(p_-;x^\p,x^+), a_{l',\p}^{b+}(p_-';x'^\p,x^+)\Big]=\de_{ll'}
\de^{ab}\de(p_--p_-')\de(x^\p-x'^\p),
\nom}
\disn{27}{
\Big[b_{l,\p}^{a}(p_-;x^\p,x^+), b_{l',\p}^{b+}(p_-';x'^\p,x^+)\Big]=-\de_{ll'}
\de^{ab}\de(p_--p_-')\de(x^\p-x'^\p),
\nom}
and  the other commutators are equal to zero.

The free (quadratic in fields) part of the Hamiltonian takes the following form in terms of these operators:
\disn{28}{
H=\int dx^\p \int_\e^\infty dp_- \sum_{l=1}^3 \frac{r_l}{2p_-}\Biggl[-m_l^2a_{l}^{a+}(p_-;x^\p,x^+)
a_{l}^{a}(p_-;x^\p,x^+)-\ns-(\partial_\p-m)a_{l}^{a+}(p_-;x^\p,x^+)
(\partial_\p-m)a_{l}^{a}(p_-;x^\p,x^+)+\ns
+m_l\ls(\partial_\p-m)
a_{l}^{a+}(p_-;x^\p,x^+)\ls a_{l,\p}^{a}(p_-;x^\p,x^+)+
b_{l,\p}^{a}(p_-;x^\p,x^+)\rs+\text{h.c.}\rs-\ns
-m_l^2\ls\frac{1-r_l}{2}
a_{l,\p}^{a+}(p_-;x^\p,x^+)+\frac{1+r_l}{2}
b_{l,\p}^{a+}(p_-;x^\p,x^+)\rs\times\ns\times
\ls\frac{1-r_l}{2}
a_{l,\p}^{a}(p_-;x^\p,x^+)+\frac{1+r_l}{2}
b_{l,\p}^{a}(p_-;x^\p,x^+)\rs+\ns+m_l\Bigg((\partial_\p-m)
a_{l}^{a+}(p_-;x^\p,x^+)\ls\frac{1-r_l}{2}
a_{l,\p}^{a}(p_-;x^\p,x^+)+\frac{1+r_l}{2}
b_{l,\p}^{a}(p_-;x^\p,x^+)\rs+\text{h.c.}\Biggr)\Biggr].
\nom}
The introduction of new variables analogous to the creation and annihilation operators gives rather simple dependence of the Hamiltonian on the $m_l$. However the quadratic in fields part of the Hamiltonian has  no diagonal form yet. It can be written, in a principal,  in the completely diagonal form but the obtained at that expression has rather cumbersome form.

\section{Conclusion}
In the present paper we consider the construction of the renormalized LF Hamiltonian
for SU(N)-gauge invariant Yang-Mills theory in (2+1)-dimensional space-time. In this case the perturbation theory in coupling constant  is superrenormalizable and only finite number of UV divergent  diagrams should be renormalized.
This gives a possibility to find the counterterms  renormalizing the theory explicitly in terms of original parameters. On the other side the restoration of the perturbative (to all orders in coupling constant) equivalence between the obtained LF theory and usual one,  quantized on the instant time surface in Lorentz coordinates, has  required the introduction of extra fields analogous to those used in the Pauli-Villars regularization.
These fields generate the Fock space on the LF with indefinite metric. Nonperturbative calculations with this Hamiltonian are very complicated \cite{Malyshev.Paston.Prokhvatilov.Zubov.IntJTheorPhys2015,
Brodsky1,Brodsky2,Brodsky3,Brodsky4}.
In the paper \cite{Malyshev.Paston.Prokhvatilov.Zubov.Franke.arXiv.2015.AIP} we have considered the simplest example: the calculation of the spectrum of the anharmonic oscillator modified by the addition of extra variables analogous to extra fields of  Pauli-Villars regularization.
In this case one can well see how the part of the spectrum related to these extra variables become separated from the usual spectrum of the model and tends to infinite values in the limit corresponding to the removing of  Pauli-Villars regularization.

As an alternative approach to the construction of the renormalized LF Hamiltonian we can mention the approach described in papers \cite{Glazek.Wilson.Phys.Rev.D1993,Glazek.Wilson.Phys.Rev.D1994} and in paper \cite{Stanislaw.Glazek.and.Maria.Gomez.Rocha.arXiv.2015} (which contains all  references to the previous papers).
This approach applies the method of approximate diagonalization of the Hamiltonian with the help of  unitary transformation depending on the scale parameter.
Such transformation is to be built using the perturbative theory in coupling constant.
In this way the approximate expression for the renormalized LF Hamiltonian can be written in terms of effective "composite particle" creation and annihilation operators.
However that procedure is rather complicated and limited by only lowest perturbation theory orders.

\section*{Appendix. Calculation of diagrams contributing to the counterterm in the renormalized LF Hamiltonian.}
Accordingly to eq. (\ref{15}) the propagator is zero if at least one of its Lorentz indices is "$-$"{}, i.e. $\De_{\m-}^{ab}=0$. Considering    the other Lorentz indices only  one can write the propagator  in the form of the following matrix:
\disn{29}{
\Delta^{ab}(k)=\begin{pmatrix}
\De_{++}^{ab}(k) & \De_{+\p}^{ab}(k) \\
\De_{\p+}^{ab}(k) & \De_{\p\p}^{ab}(k)
\end{pmatrix}=\frac{i\de^{ab}}{k^2-m^2+i0}
\begin{pmatrix}
2k_+ & k_\p-im \\
k_\p+im & k_-
\end{pmatrix}.
\nom}
The following factor corresponds to the vertex:
\disn{30}{
V^{\m\n\rho,abc}(p,k,q)=-gf^{abc}\ls(p-k)^\rho g^{\m\n}+(k-q)^\m g^{\n\rho}+(q-p)^\n g^{\m\rho}\rs,
\nom}
where the metric tensor $g^{\m\n}$ has the following nonzero components: $g^{+-}=g^{-+}=1$, $g^{\p\p}=-1$.
Since propagators are connected to vertices in diagrams for Green functions it is not necessary to consider vertices with  Lorentz indices "$-$"{} (let us remark that this would be necessary if  we use Ward identities explicitly).
We can form the following matrix from the vertex  one of Lorentz indices of which is fixed:
\disn{31}{
V^{\m,abc}(p,k,q)=\begin{pmatrix}
V^{\m++,abc}(p,k,q) & V^{\m+\p,abc}(p,k,q) \\
V^{\m\p+,abc}(p,k,q) & V^{\m\p\p,abc}(p,k,q)
\end{pmatrix}=\ns=-gf^{abc}\begin{pmatrix}
0 & (q-p)_-g^{\m\p} \\
(p-k)_- g^{\m\p} & (k-q)_\p g^{\m\p}+(q-k)^\m
\end{pmatrix}.
\nom}
We obtain the following expression for the diagram shown in Fig.~\ref{Fig3}a:
$$D^{\m_1\m_2,a_1a_2}(k)=\int dq_+dq_-dq_\p
\ls
V^{\m_1\n_1\rho_1,a_1b_1c_1}(k,q)\,
\Delta_{\rho_1\n_2}^{c_1b_2}(k+q)\, V^{\m_2\n_2\rho_2,a_2b_2c_2}(k,q)\,\Delta_{\rho_2\n_1}^{c_2b_1}(q)\rs=$$
\disn{a1}{=
-g^2N\de^{a_1a_2}\int \frac{dq_+dq_-dq_\p}{(q^2-m^2+i0)((k+q)^2-m^2+i0)}\times\ns\times Tr\Bigg(
\begin{pmatrix}
0 & (q_- + 2 k_-) g^{\m_1\p} \\
(q_--k_-) g^{\m_1\p} & 2 q^{\m_1} + k^{\m_1}-(k_\p + 2 q_\p)g^{\m_1\p}
\end{pmatrix}
\begin{pmatrix}
2(k_++q_+) & k_\p+q_\p-im \\
k_\p+q_\p+im & k_--q_-
\end{pmatrix}\times\ns\times
\begin{pmatrix}
0 & (q_--k_-) g^{\m_2\p} \\
(q_- + 2 k_-) g^{\m_2\p} & 2 q^{\m_2} + k^{\m_2}-(k_\p + 2 q_\p) g^{\m_2\p}
\end{pmatrix}
\begin{pmatrix}
2q_+ & q_\p-im \\
q_\p+im & q_-
\end{pmatrix}\Bigg).
\nom}
We used here the following equality: $f^{a_1b_1b_2}f^{a_2b_1b_2}=N\de^{a_1a_2}$.
To calculate the counterterm we consider the difference between the expressions for this integral in our and in dimensional regularization,  taking into account the first two terms of the Tailor series in $k$ in the vicinity of the point $k=0$.
Besides we do the Euclidean  rotation in the momentum space.  Then one can see easily that all components of the diagram containing Lorentz index "+"{} disappear.
The remaining component with the indices $\p\p$ can be calculated analytically and gives the finite value for it in the UV regularization removing limit.
The analogous expression can be written for the diagram shown in Fig.~\ref{Fig3}d. We calculate it at zero external momenta because this diagram  can diverge only logarithmically.
Despite of its  complicated expression this diagram allows analytical consideration after the Euclidean rotation in the momentum space.  In result we obtain that this diagram is equal to zero at zero external momenta.
The expressions for the remaining logarithmically divergent diagrams \ref{Fig3}b, \ref{Fig3}c,  having Lorentz indices "$\p\p$"{},  can be obtained in analogous way using the formulas (\ref{29}), (\ref{31}). However,  these expressions turn out to be much more complicated.
Besides, we were not able to consider them analytically. Only approximate numerical calculations are possible, and its results can be used,  nevertheless, in nonperturbative Hamiltonian approach to this model.

{\bf Acknowledgements.}
Authors thank the organizers of the International Conference  "Models of Quantum Field Theory" (MQFT-2015)  dedicated to the 75 years of birthday of A.N.~Vasil'ev. Also they thank S.A.~Paston and M.V.~Kompaniets for useful discussions.  The work of M.Yu.~Malyshev,  E.V.~Prokhvatilov and R.A.~Zubov was supported by the grant of SPbGU  N~11.38.189.2014.

%\bibliographystyle{my2beznazv}
%\bibliography{Malyshev_English}
%\bibliography{Malyshev_Russian,Malyshev_English}

\begin{thebibliography}{10}
\newcommand{\enquote}[1]{``#1''}
\providecommand{\url}[1]{\texttt{#1}}
\providecommand{\urlprefix}{URL }
\providecommand{\eprint}[2][]{\url{#2}}

\bibitem{tmf97}
V.~A. Franke, S.~A. Paston, \emph{Theor.~Math.~Phys.}, \textbf{112}:3 (1997),
  1117--1130, arXiv:hep-th/9901110.

\bibitem{tmf99}
S.~A. Paston, E.~V. Prokhvatilov, V.~A. Franke, \emph{Theor.~Math.~Phys.},
  \textbf{120}:3 (1999), 1164--1181, arXiv:hep-th/0002062.

\bibitem{tmf02}
S.~A. Paston, E.~V. Prokhvatilov, V.~A. Franke, \emph{Theor.~Math.~Phys.},
  \textbf{131}:1 (2002), 516--526, arXiv:hep-th/0302016.

\bibitem{Paston.Prokhvatilov.Franke.Nucl.Phys.B.Proc.Suppl.2002}
S.~A. Paston, E.~V. Prokhvatilov, V.~A. Franke,
  \emph{Nucl.~Phys.~B~(Proc.~Suppl.)}, \textbf{108} (2002), 189--193,
  arXiv:hep-th/0111009.

\bibitem{Brodsky.Franke.Hiller.McCartor.Paston.Prokhvatilov2004}
S.~J. Brodsky, V.~A. Franke, J.~R. Hiller, G.~McCartor, S.~A. Paston, E.~V.
  Prokhvatilov, \emph{Nucl.~Phys.~B}, \textbf{703} (2004), 333--362,
  arXiv:hep-ph/0406325.

\bibitem{Yad.Fiz.2005}
S.~A. Paston, E.~V. Prokhvatilov, V.~A. Franke, \emph{Phys.~Atom.~Nucl.},
  \textbf{68} (2005), 267--278, arXiv:hep-th/0501186.

\bibitem{NPPF}
V.~A. Franke, {Yu. V. Novozhilov}, S.~A. Paston, E.~V. Prokhvatilov, Focus on
  quantum field theory, chap. Quantization of Field Theory on the Light Front,
  23--81, Nova science publishers, New York, 2005, arXiv:hep-th/0404031.

\bibitem{Malyshev.Paston.Prokhvatilov.Zubov.IntJTheorPhys2015}
M.~{Yu.} Malyshev, S.~A. Paston, E.~V. Prokhvatilov, R.~A. Zubov,
  \emph{Int.~J.~Theor.~Phys.}, \textbf{54} (2015), 169--184, arXiv:1311.4381
  [hep-th].

\bibitem{Malyshev.Paston.Prokhvatilov.Zubov.Franke.TMPh2015}
M.~{Yu.} Malyshev, S.~A. Paston, E.~V. Prokhvatilov, R.~A. Zubov, V.~A. Franke,
  \emph{Theor.~Math.~Phys.}, \textbf{184}:3 (2015), 1314--1323,
  arXiv:1505.00272 [hep-th].

\bibitem{dir}
P.~A.~M. Dirac, \emph{Rev.~Mod.~Phys.}, \textbf{21}:3 (1949), 392--398.

\bibitem{burlang}
M.~Burkardt, A.~Langnau, \emph{Phys.~Rev.~D}, \textbf{44}:4 (1991), 1187--1197.

\bibitem{Burkardt.Langnau}
M.~Burkardt, A.~Langnau, \emph{Phys.~Rev.~D}, \textbf{44}:12 (1991),
  3857--3867.

\bibitem{Prokhvatilov.Franke.Phys.Atom.Nucl.1989}
E.~V. Prokhvatilov, V.~A. Franke, \emph{Sov. J. Nucl. Phys.}, \textbf{49}
  (1989), 688.

\bibitem{Ilgen}
E.M. Ilgenfritz, V.~A. Franke, S.~A. Paston, H.~J. Pirner, E.~V. Prokhvatilov,
  \emph{Theor.~Math.~Phys.}, \textbf{148}:1 (2006), 948--959,
  arXiv:hep-th/0610020.

\bibitem{Stanislaw.Glazek.and.Maria.Gomez.Rocha.arXiv.2015}
S.~D. Glazek, M.~G\'{o}mez-Rocha, \enquote{Asymptotic freedom of gluons in the
  Fock space}, 2015, arXiv:1510.01609 [hep-th].

\bibitem{Glazek.Wilson.Phys.Rev.D1993}
S.~D. Glazek, K.~G. Wilson, \emph{Phys.~Rev.~D}, \textbf{48}:8 (1993),
  4214--4218.

\bibitem{Glazek.Wilson.Phys.Rev.D1994}
S.~D. Glazek, K.~G. Wilson, \emph{Phys.~Rev.~D}, \textbf{49}:12 (1994),
  5863--5872.

\bibitem{PV}
W.~Pauli, F.~Villars, \emph{Rev.~Mod.~Phys.}, \textbf{21}:3 (1949), 434--444.

\bibitem{Deser.Jackiw.Templeton.Annals.of.Physics.2000}
S.~Deser, R.~Jackiw, S.~Templeton, \emph{Annals of Physics}, \textbf{281}
  (2000), 409--449.

\bibitem{Deser.Jackiw.Templeton.Phys.Rev.Lett.1982}
R.~Jackiw, S.~Templeton, \emph{Phys.~Rev.~Lett.}, \textbf{48} (1982), 975--978.

\bibitem{Brodsky1}
S.~J. Brodsky, J.~R. Hiller, G.~McCartor, \emph{Phys.~Rev.~D}, \textbf{64}:11,
  114023, arXiv:hep-ph/0107038.

\bibitem{Brodsky2}
S.~J. Brodsky, J.~R. Hiller, G.~McCartor, \emph{Ann.~Phys.}, \textbf{296}:2
  (2002), 406--424, arXiv:hep-th/0107246.

\bibitem{Brodsky3}
S.~J. Brodsky, J.~R. Hiller, G.~McCartor, \emph{Phys.~Rev.~D}, \textbf{60}:5,
  054506, arXiv:hep-ph/9903388.

\bibitem{Brodsky4}
S.~J. Brodsky, J.~R. Hiller, G.~McCartor, \emph{Phys.~Rev.~D}, \textbf{58}:2,
  025005, arXiv:hep-th/9802120.

\bibitem{Malyshev.Paston.Prokhvatilov.Zubov.Franke.arXiv.2015.AIP}
M.~{Yu.} Malyshev, S.~A. Paston, E.~V. Prokhvatilov, R.~A. Zubov, V.~A. Franke,
  \enquote{Pauli-Villars Regularization in nonperturbative Hamiltonian approach
  on the Light Front}, 2015, arXiv:1504.07951 [hep-th].

\end{thebibliography}

\end{document}